\documentclass[showpacs,showkeys,preprintnumbers]{revtex4}%
\usepackage{amsfonts}
\usepackage{amsmath}
\usepackage{graphicx}
\usepackage{dcolumn}
\usepackage{bm}
\usepackage{amssymb}%
\newcommand\T{\rule{0pt}{3.7ex}}
\newcommand\B{\rule[-3ex]{0pt}{0pt}}
\begin{document}
\title{Decays of the vector glueball }
\author{Francesco Giacosa$^{\text{(a,b)}}$}
\email{fgiacosa@ujk.edu.pl}
\author{Julia Sammet$^{\text{(b)}}$ }
\email{sammet@th.physik.uni-frankfurt.de}
\author{Stanislaus Janowski$^{\text{(b)}}$ }
\email{janowski@th.physik.uni-frankfurt.de}
\affiliation{$^{\text{(a)}}$ Institute of Physics, Jan Kochanowski University, ul.
Swietokrzyska 15, 25-406 Kielce, Poland,}
\affiliation{$^{\text{(b)}}$Institute for Theoretical Physics, Goethe University,
Max-von-Laue-Str.\ 1, 60438 Frankfurt am Main, Germany.}

\begin{abstract}
We calculate two- and three-body decays of the (lightest) vector glueball into
(pseudo)scalar, (axial-)vector, as well as pseudovector and excited vector
mesons\textbf{ }in the framework of a model of QCD. While absolute values of
widths cannot be predicted because the corresponding coupling constants are
unknown, some interesting branching ratios can be evaluated\textbf{ }by
setting the mass of the yet hypothetical vector glueball to $3.8$ GeV as
predicted by quenched Lattice QCD\textbf{.} We find that the decay mode
$\omega\pi\pi$ should be one of the largest (both through the decay chain
$\mathcal{O}\rightarrow b_{1}\pi\rightarrow$ $\omega\pi\pi$ and through the
direct coupling $\mathcal{O}\rightarrow\omega\pi\pi$)$.$ Similarly, the
(direct and indirect) decay into $\pi KK^{\ast}(892)$ is sizable. Moreover,
the decays into $\rho\pi$ and $K^{\ast}(892)K$ are, although subleading,
possible and could play a role in explaining the $\rho\pi$ puzzle of the
charmonium state $\psi(2S)$ thank to a (small) mixing with the vector
glueball. The vector glueball can be directly formed at the ongoing BESIII
experiment as well as at the future PANDA experiment at the FAIR facility. If
the width is sufficiently small ($\lesssim100$ MeV) it should not escape
future detection. It should be stressed that the employed model is based on
some inputs and simplifying assumptions: the value of glueball mass (at
present, the quenched lattice value is used), the lack of mixing of the
glueball with other quarkonium states, and the use of few interaction terms.
It then represents a first step toward the identification of the main decay
channels of the vector glueball, but shall be improved when corresponding
experimental candidates and/or new lattice results will be available.

\end{abstract}

\maketitle

\section{Introduction}

The search for glueballs is an important part of experimental as well as
theoretical hadronic physics\ \cite{review,revglueball}. Lattice QCD predicts
a rich spectrum of glueball states below 5 GeV \cite{mainlattice,morningstar},
but up to now no predominantly glueball state could be unambiguously assigned
to one of the mesons listed in the PDG \cite{pdg}.

The lightest glueball predicted by lattice QCD is a scalar particle. This is
definitely the glueball which has been most intensively studied both
experimentally and theoretically. Various theoretical approaches were
developed to understand which scalar-isoscalar resonance contains the largest
gluonic amount, e.g. Refs. \cite{revglueball,close} and refs. therein. In most
scenarios, either the resonance $f_{0}(1500)$ or the resonance $f_{0}(1710)$
has the largest gluonic content in its wave function.

The field corresponding to the scalar glueball is often related to the dilaton
\cite{migdal,schechter}: the corresponding nonvanishing condensate is
proportional to the gluon condensate and is linked to a basic feature of QCD,
the trace anomaly, which allows to understand how a dimension enters into a
classically dimensionless theory. Indeed, dilatation invariance and its
anomalous breaking in the Yang-Mills sector have been, together with chiral
symmetry, the guiding principle behind the development of an effective
hadronic model of QCD, the so-called extended Linear Sigma Model (eLSM)
\cite{dick,staninew}. The eLSM has shown to be capable to describe various
hadronic masses and decays below 1.8 GeV, as the fit in Ref. \cite{dick}
confirms. The eLSM can be coupled to glueballs allowing to calculate their
decays, see the following discussion. The scalar glueball is automatically
present in the eLSM as a dilaton and is coupled to light mesons: its
assignment and its mixing with quark-antiquark scalar fields have been studied
in\ Ref. \cite{staninew}, where only one assignment was found to be
acceptable: $f_{0}(1710)$ is mostly gluonic. Besides the eLSM, evidence that
$f_{0}(1710)$ has a large gluonic amount is recently mounting from both
lattice studies \cite{chenlattice} and from an holographic approach
\cite{rebhan}.

The pseudoscalar glueball has also received much attention (see Ref.
\cite{masoni} for a review), since it is linked to another important feature
of QCD: the chiral anomaly \cite{schechterpsg}. Lattice QCD predicts a mass of
about $2.6$ GeV, but the scenario in which the pseudoscalar glueball is light
(at about $1.5$ GeV) was also widely investigated, e.g. Ref.
\cite{masoni,tichy} and refs. therein. Yet, the mismatch with the mass
calculated from Lattice QCD seems to be too large for this scenario to be
realistic. In a recent work, the branching ratios of a putative pseudoscalar
glueball with a mass of about $2.6$ GeV were calculated within the eLSM
mentioned above \cite{psg,psgproc}. To this end, a chirally invariant
interaction term coupling the pseudoscalar glueball to light mesons was
considered. It was found that the channel $\pi\pi K$ is dominant and that the
channels $\pi\pi\eta$ and $\pi\pi\eta^{\prime}$ are sizable. Such informations
can be useful in future experimental search of the pseudoscalar glueball at
PANDA \cite{panda} and at BESIII \cite{bes}. (Very recently, the very same
interaction Lagrangian has been used to study the branching ratios of a second
excited pseudoscalar glueball with a mass of about $3.7$ GeV \cite{walaaepsg}.)

In this work we continue the investigation of glueballs by concentrating our
attention to the vector glueball, denoted as $\mathcal{O}$. The vector
glueball has received some attention in the past (e.g. in Refs.
\cite{brodsky,hou,suzuki,robson,besvg} and refs. therein, where it was studied
in the context of the so-called $\rho\pi$ puzzle, see also below), but to our
knowledge a systematic prediction of its decay rates has not yet been
performed. Our plan is to study the vector glueball in the framework of the
mentioned model of QCD, the eLSM, properly coupled to the vector glueball. As
a consequence, the results are model dependent and only branching ratios can
be calculated (full decay rates depend on coupling constant which cannot be
calculated). Also the mass of the glueball cannot be determined within the
model. We thus use the value predicted by quenched lattice QCD ($3.8$ GeV)
\cite{mainlattice}, well above the scalar and pseudoscalar ones [The
discussion concerning the use of the lattice mass as an input and the validity
of our interaction terms for the evaluation of the glueball's decays are
presented in Sec. IV]. It is then expected that the vector glueball decays in
two and three light mesons. We thus plan to evaluate, in the framework of our
model, various branching ratios.\ 

In order to obtain the interaction Lagrangians, we need to couple the vector
glueball to conventional light mesons contained in the eLSM. Being glueballs
chirally invariant fields, we \textbf{assume that} the interactions fulfill
chiral invariance (and, at least for the dominant terms, also dilatation
invariance). Moreover, in addition to the standard pseudoscalar, scalar,
vector, and axial-vector mesons studied in Ref. \cite{dick}, we will also
consider pseudovector and excited vector mesons. It is important to stress
from the very beginning that our approach could not yet be tested
experimentally. (In particular, for what concerns heavy glueballs, there are
at present no known candidates). Moreover, we shall also neglect mixing of the
glueball with conventional quarkonium states. As we discuss more in detail
later on, it will be possible to model in the future such mixing when more
information will be available.

With these cautionary remarks in mind, we aim to make some observations
concerning the decays of the vector glueball. In particular, we will show
that, in our calculational approach, one of the most important decay modes of
the vector glueball is the decay into a pseudoscalar-pseudovector pair (in
particular, the channel $\mathcal{O}\rightarrow b_{1}\pi\rightarrow\omega
\pi\pi$). These decays are obtained from an interaction term (our first
effective Lagrangian for $\mathcal{O}$, denoted as $\mathcal{L}_{1}$) which
involves the least possible number of quark-antiquark fields (only two; hence,
to assure dilatation invariance, also the dilaton field is coupled). As a side
effect, this analysis also offers the mathematical basis for an extension of
the eLSM containing pseudovector and excited vector fields.

The second chiral Lagrangian that we build is denoted as $\mathcal{L}_{2}$ and
contains the coupling of $\mathcal{O}$ to (pseudo)scalar and (axial-)vector
fields. Here, three quark-antiquark fields are at first present at each
vertex, hence three-body decay of \ $\mathcal{O}$ are automatically obtained.
The decays of $\mathcal{O}$ into two pseudoscalar mesons and one vector mesons
are dominant. In particular, the direct decay channels $\mathcal{O}%
\rightarrow\omega\pi\pi$ and $\mathcal{O}\rightarrow K^{\ast}(892)K\pi$ are
the largest. Both the first and the second Lagrangians predict a decay into
$\omega\pi\pi$ and into $K^{\ast}(892)K\pi$ (indirect in the first case,
direct in the second), which then represent the two golden channels toward a
possible future detection of the vector glueball. In the end, when
condensation of scalar fields is taken into account, $\mathcal{L}_{2}$ also
delivers two-body decays. Here, one vector and one scalar fields are the most
relevant decay channels.

Finally, we shall also analyze a third interaction Lagrangian (denoted as
$\mathcal{L}_{3}$), in which four quark-antiquark fields are present at each
vertex. This Lagrangian breaks dilatation invariance (the coupling constant is
proportional to Energy$^{-2}$), yet we decided to study it since it delivers
interesting decays into one vector and one axial-vector pair and into one
vector and one pseudoscalar pair. In particular, the $\rho\pi$ and $K^{\ast
}(892)K$ decay modes are relevant in connection to the so-called $\rho\pi
$-puzzle of the state $\psi(2S),$ see the discussion in Sec. IV.

The vector glueball can be constructed with (at least) three constituent
gluons,\textbf{ }hence its decays and mixing are expected to be sufficiently
small to allow detection (even if in the context of our model we cannot
calculate them). Even if the predicted mass of $\mathcal{O}$ lies at about
$3.8$\ GeV and hence above the $DD$ threshold, we do not find a sizable decay
into charmed mesons. Namely, the $DD$ channel turns out to be small (albeit
nonzero), while other channels with a potentially sizable interaction -such as
the pseudoscalar-pseudovector channel mentioned above- are not kinematically
allowed when charmed mesons are considered.

In the future, the PANDA experiment at FAIR \cite{panda} will be able to form
glueballs and to study their decays. In particular, the vector glueball has
the quantum numbers of the photon, hence it can be also directly formed at
BES\textrm{III} (for which an energy scan in the region $3.5$ to $4$ GeV with
particular attention to the $\omega\pi\pi$ and $K^{\ast}(892)K\pi$ channels
would be necessary). It should be mentioned that in a previous study at BES
\cite{besvg}, a vector glueball was searched --without success-- in the
$\rho\pi$ channel, which is however not the favorite decay mode found in our study.

This paper is organized as follows. In Sec.\ II we show the quark-antiquark
nonets and their properties, while in Sec.\ III we introduce three effective
Lagrangians and calculate decay ratios. In Sec. IV we present various
discussions: the validity and the limits of the used interaction Lagrangians,
the use of the mass from lattice QCD as an input, the large-$N_{c}$ limit, and
the so-called $\rho\pi$ puzzle of $\psi(2S).$ Finally, in Sec. V we outline
our conclusions and outlooks. In the Appendices we show relevant details of
the eLSM and of the calculations.

\section{Chiral multiplets}

In this Section, we concentrate on the quark-antiquark fields which represent
the decay products of the vector glueball. We split our study into two parts:
first, we consider (pseudo)scalar and (axial-)vector fields, which are also
the basic ingredients of the eLSM when proper chiral combinations of them are
taken into account. Then, in the second part we describe pseudovector mesons
and excited vector mesons and the corresponding chiral combination.

\subsection{(Pseudo)scalar and (axial-)vector quark-antiquark multiplets}

The nonet of pseudoscalar fields is introduced as
\begin{equation}
P=\frac{1}{\sqrt{2}}\left(
\begin{array}
[c]{ccc}%
\frac{\eta_{N}+\pi^{0}}{\sqrt{2}} & \pi^{+} & K^{+}\\
\pi^{-} & \frac{\eta_{N}-\pi^{0}}{\sqrt{2}} & K^{0}\\
K^{-} & \bar{K}^{0} & \eta_{S}%
\end{array}
\right)  \text{ ,}%
\end{equation}
which contains the renowned light pseudoscalar nonet $\{\pi$, $K,\eta
,\eta^{\prime}\}$ \cite{pdg}, where $\eta$ and $\eta^{\prime}$ arise via the
mixing $\eta=\eta_{N}\cos\theta_{p}+\eta_{S}\sin\theta_{p},$ $\eta^{\prime
}=-\eta_{N}\sin\theta_{p}+\eta_{S}\cos\theta_{p}$ with $\theta_{p}%
\simeq-44.6^{\circ}$ \cite{dick}. (Using other values as e.g. $\theta
_{p}=-41.4^{\circ}$ \cite{kloe2} changes only slightly the results presented
in this work). As a next step, we introduce the matrix of scalar fields
\begin{equation}
S=\frac{1}{\sqrt{2}}\left(
\begin{array}
[c]{ccc}%
\frac{\sigma_{N}+a_{0}^{0}}{\sqrt{2}} & a_{0}^{+} & K_{S}^{+}\\
a_{0}^{-} & \frac{\sigma_{N}-a_{0}^{0}}{\sqrt{2}} & K_{S}^{0}\\
K_{S}^{-} & \bar{K}_{S}^{0} & \sigma_{S}%
\end{array}
\right)  \text{ ,}%
\end{equation}
that contains the fields $\{a_{0}(1450),K_{0}^{\ast}(1430),\sigma_{N}%
,\sigma_{S}\}.$ (Note, the scalar quark-antiquark states lie above $1$ GeV
\cite{dick}, thus the nonet of light scalar states $\{a_{0}(980),K_{0}^{\ast
}(800),f_{0}(500),f_{0}(980)\}$ is something else. A possibility is a nonet of
light tetraquark states \cite{jaffe,varietq} or a nonet of dynamically
generated states \cite{lowscalars,sigmareview,thomas}). As a first
approximation, the non-strange bare field $\sigma_{N}\equiv\left\vert \bar
{u}u+\bar{d}d\right\rangle /\sqrt{2}$ corresponds predominantly to the
resonance \thinspace$f_{0}(1370)$ and the bare field $\sigma_{S}%
\equiv\left\vert \bar{s}s\right\rangle $ predominantly to $f_{0}(1500).$
Finally, in the eLSM the state $f_{0}(1710)$ is predominantly a scalar
glueball, see details in\ Ref. \cite{staninew} in which the mixing matrix is presented.

Nowadays evidence toward a large gluonic amount in $f_{0}(1710)$ is
increasing: besides the eLSM\ \cite{staninew}, in the recent lattice work of
Ref. \cite{chenlattice} the radiative decay $j/\psi\rightarrow\gamma G$ (where
$G$ is a pure glueball) has been analyzed and found to be in good agreement
with the experimental decay rate $j/\psi\rightarrow\gamma f_{0}(1710).$
Moreover, the study of Ref. \cite{rebhan} reaches the same conclusion in the
framework of holographic QCD.

The scalar and pseudoscalar matrices are combined into%

\begin{equation}
\Phi=S+iP\text{ ,} \label{phimat}%
\end{equation}
which has a simple transformation under chiral transformations $U_{L}(3)\times
U_{R}(3)$: $\Phi\rightarrow U_{L}\Phi U_{R}^{\dagger}$, where $U_{L}$ and
$U_{R}$ are $U(3)$ matrices. Under parity $\Phi\rightarrow\Phi^{\dagger}$ and
under charge conjugation $\Phi\rightarrow\Phi^{t}.$ The matrix $\Phi$ is used
as a building block in the construction of the eLSM Lagrangian, see Appendix A
and Tables I and II.

We now turn to vector and axial-vector fields. The nonet of vector fields is
introduced as
\begin{equation}
V^{\mu}=\frac{1}{\sqrt{2}}\left(
\begin{array}
[c]{ccc}%
\frac{\omega_{N}+\rho^{0}}{\sqrt{2}} & \rho^{+} & K^{\star+}\\
\rho^{\mu-} & \frac{\omega_{N}-\rho^{0}}{\sqrt{2}} & K^{\star0}\\
K^{\star-} & \bar{K}^{\star0} & \omega_{S}%
\end{array}
\right)  ^{\mu}\;\text{,}%
\end{equation}
and the nonet of axial-vector fields as
\begin{equation}
A^{\mu}=\frac{1}{\sqrt{2}}\left(
\begin{array}
[c]{ccc}%
\frac{f_{1N}+a_{1}^{0}}{\sqrt{2}} & a_{1}^{+} & K_{1,A}^{+}\\
a_{1}^{-} & \frac{f_{1N}-a_{1}^{0}}{\sqrt{2}} & K_{1,A}^{0}\\
K_{1,A}^{-} & \bar{K}_{1,A}^{0} & f_{1S}%
\end{array}
\right)  ^{\mu}\;\text{.} \label{noneta}%
\end{equation}
The matrix $V^{\mu}$ contains the vector states $\{\rho,K^{\ast}%
(892),\omega,\phi\}$, while the matrix $A^{\mu}$ contains the axial-vector
states $\{a_{1}(1230),$ $K_{1,A},$ $f_{1}(1285),$ $f_{1}(1420)\}$ , where
$K_{1,A}$ is a mixture of the two physical states $K_{1}(1270)$ and
$K_{1}(1400),$ see Sec. II.B. We neglect (the anyhow small) strange-nonstrange
mixing, hence $\omega\equiv\omega_{N}$ and $f_{1N}\equiv f_{1}(1285)$ are
purely nonstrange mesons of the type $\sqrt{1/2}(\bar{u}u+\bar{d}d)$, while
$\omega_{S}\equiv\phi$ and $f_{1S}\equiv f_{1}(1285)$ are purely $\bar{s}s$ states.

Then, one defines the right-handed and left-handed combinations:
\begin{equation}
R^{\mu}=V^{\mu}-A^{\mu}\text{ and }L^{\mu}=V^{\mu}+A^{\mu}\text{ .} \label{rl}%
\end{equation}
Under chiral transformation they transform in a compact way: $R^{\mu
}\rightarrow U_{R}R^{\mu}U_{R}^{\dagger}$, $L^{\mu}\rightarrow U_{L}L^{\mu
}U_{L}^{\dagger}.$\newline Details of the currents and transformations are
shown in Table I and II.

The eLSM Lagrangian includes the multiplets $S,$ $P,$ $V,$ and $A$ presented
above. In addition, a dilaton/glueball field is also present in order to
describe dilatation symmetry and its anomalous breaking. The details of the
eLSM (together with its symmetries, most notably chiral and dilatation
symmetries together with their anomalous, explicit, and spontaneous breaking
terms) are briefly summarized in\ Appendix A and extensively presented in
Refs. \cite{dick,staninew} for $N_{f}=3.$ (Previous versions of the eLSM for
$N_{f}=2$ are discussed in\ Ref. \cite{denisnf2} while an extension to
$N_{f}=4$ can be found\ in Ref. \cite{walaa}. Baryons are considered in Ref.
\cite{gallas}, while properties at finite density are studied in Ref.
\cite{giuseppe} and at finite temperature in\ Ref. \cite{kovacs}.\ Recently,
it was also shown that the low-energy limit of the eLSM is equivalent to
chiral perturbation theory \cite{elsmchpt}.)

\subsection{Pseudovector and excited vector mesons}

We aim to investigate also the decay of the vector glueball into pseudovector
and excited vector mesons. To this end we introduce the matrix%

\begin{equation}
B^{\mu}=\frac{1}{\sqrt{2}}%
\begin{pmatrix}
\frac{h_{1,N}+b_{1}^{0}}{\sqrt{2}} & b_{1}^{+} & K_{1,B}^{\star+}\\
b_{1}^{-} & \frac{h_{1,N}+b_{1}^{0}}{\sqrt{2}} & K_{1,B}^{\star0}\\
K_{1,B}^{\star-} & \bar{K}_{1,B}^{\star0} & h_{1,S}%
\end{pmatrix}
^{\mu} \label{nonetb}%
\end{equation}
that describes the nonet of pseudovector resonances
\begin{equation}
\{b_{1}(1230),K_{1,B},h_{1}(1170),h_{1}(1380)\}\text{ .}%
\end{equation}
Also here, the strange-nonstrange isoscalar mixing is neglected, thus
$h_{1,N}\equiv h_{1}(1170)$ is a purely nonstrange state, while $h_{1,S}\equiv
h_{1}(1380)$ is a purely strange-antistrange state.

The kaonic fields $K_{1,A}$ from Eq. (\ref{noneta}) and $K_{1,B}$ from Eq.
(\ref{nonetb}) mix and generate the two physical resonances $K_{1}(1270)$ and
$K_{1}(1400)$:%
\begin{equation}%
\begin{pmatrix}
K_{1}^{+}(1270)\\
K_{1}^{+}(1400)
\end{pmatrix}
^{\mu}=%
\begin{pmatrix}
\cos\varphi & -i\sin\varphi\\
-i\sin\varphi & \cos\varphi
\end{pmatrix}%
\begin{pmatrix}
K_{1,A}^{+}\\
K_{1,B}^{+}%
\end{pmatrix}
^{\mu}\text{ .} \label{mixk}%
\end{equation}
The mixing angle reads $\varphi=(56.3\pm4.2)^{\circ}$ \cite{florianlisa}. The
same transformations hold for $K_{1}^{0}(1270)$ and $K_{1}^{0}(1400),$ while
for the other kaonic states one has to take into account that $K_{1}%
^{-}(1270)=K_{1}^{+}(1270)^{\dagger}$ and $\bar{K}_{1}^{0}(1270)=K_{1}%
^{0}(1270)^{\dagger}$ (and so for $K_{1}^{-}(1400)$).

The chiral partners of the pseudovector mesons are excited vector mesons which
arise from the combination $L=2,$ $S=1$ coupled to $J^{PC}=1^{--}$. The
corresponding fields listed in the PDG \cite{pdg} are given by
\[
\{\rho(1700),K^{\ast}(1680),\omega(1650),\phi(???)\}\text{ .}%
\]
Note, $\phi(???)$ was not yet found (one expects a vector state with a mass of
about $1.95$ GeV from the comparison to the radially excited vector mesons).
The corresponding nonet reads:
\begin{equation}
E_{\text{ang}}^{\mu}=\frac{1}{\sqrt{2}}%
\begin{pmatrix}
\frac{\omega_{\text{ang},N}+\rho_{\text{ang}}^{0}}{\sqrt{2}} & \rho
_{\text{ang}}^{+} & K_{\text{ang}}^{\star+}\\
\rho_{\text{ang}}^{-} & \frac{\omega_{\text{ang},N}-\rho_{\text{ang}}^{0}%
}{\sqrt{2}} & K_{\text{ang}}^{\star0}\\
K_{\text{ang}}^{\star-} & \bar{K}_{\text{ang}}^{\star0} & \omega
_{\text{ang},S}%
\end{pmatrix}
^{\mu}\text{ .}%
\end{equation}
We then build the matrix%

\begin{equation}
\tilde{\Phi}^{\mu}=E_{\text{ang}}^{\mu}-iB^{\mu}\text{ ,}%
\end{equation}
which under chiral transformations changes as $\tilde{\Phi}^{\mu}\rightarrow
U_{L}\tilde{\Phi}^{\mu}U_{R}^{\dag}$ (just as the standard (pseudo)scalar
$\Phi,$ hence the name), under parity as $\tilde{\Phi}^{\mu}\rightarrow
\tilde{\Phi}^{\dag\mu}$, and under charge conjugations as $\tilde{\Phi}^{\mu
}\rightarrow-\tilde{\Phi}^{t,\mu},$ see Table I and II for details. Although
not the goal of the present paper, one could use the matrix $\tilde{\Phi}%
^{\mu}$ in order to build an extension of the eLSM which includes pseudovector
and excited vector mesons. This project is left for the future.

Finally, we present in the Tables I and II the summary of all relevant
properties and transformations of the nonets introduced in this Section.%

\begin{table}[h] \centering
\begin{tabular}
[c]{|c|c|c|c|c|c|}\hline
Nonet & $L$ & $S$ & $J^{PC}$ & Current & Assignment\rule{0pt}{3.7ex}%
\rule[-2ex]{0pt}{0pt}\\\hline
$P$ & $0$ & $0$ & $0^{-+}$ & $P_{ij}=\frac{1}{\sqrt{2}}\bar{q}_{j}i\gamma
^{5}q_{i}$ & \rule{0pt}{3.7ex}\rule[-2ex]{0pt}{0pt} $\pi,K,\eta,\eta^{\prime}%
$\\\hline
$S$ & $1$ & $1$ & $0^{++}$ & $S_{ij}=\frac{1}{\sqrt{2}}\bar{q}_{j}q_{i}$ &
$a_{0}(1450),$ $K_{0}^{\ast}(1430),$ $f_{0}(1370),$ $f_{0}(1510)$
\rule{0pt}{3.7ex}\rule[-2ex]{0pt}{0pt}\\\hline
$V^{\mu}$ & $0$ & $1$ & $1^{--}$ & $V_{ij}^{\mu}=\frac{1}{\sqrt{2}}\bar{q}%
_{j}\gamma^{\mu}q_{i}$ & $\rho(770),$ $K^{\ast}(892),$ $\omega(785)$,
$\phi(1024)$ \rule{0pt}{3.7ex}\rule[-2ex]{0pt}{0pt}\\\hline
$A^{\mu}$ & $1$ & $1$ & $1^{++}$ & $A_{ij}^{\mu}=\frac{1}{\sqrt{2}}\bar{q}%
_{j}\gamma^{5}\gamma^{\mu}q_{i}$ & $a_{1}(1230),$ $K_{1,A},$ $f_{1}(1285),$
$f_{1}(1420)$ \rule{0pt}{3.7ex}\rule[-2ex]{0pt}{0pt}\\\hline
$B^{\mu}$ & $1$ & $0$ & $1^{+-}$ & $B_{ij}^{\mu}=\frac{1}{\sqrt{2}}\bar{q}%
_{j}\gamma^{5}\overleftrightarrow{\partial}^{\mu}q_{i}$ & $b_{1}(1230),\text{
}K_{1,B},\text{ }h_{1}(1170),\text{ }h_{1}(1380)$ \rule{0pt}{3.7ex}%
\rule[-2ex]{0pt}{0pt}\\\hline
$E_{\text{ang}}^{\mu}$ & $2$ & $1$ & $1^{--}$ & $E_{\text{ang},ij}^{\mu}%
=\frac{1}{\sqrt{2}}\bar{q}_{j}i\overleftrightarrow{\partial}^{\mu}q_{i}$ &
$\rho(1700),$ $K^{\ast}(1680),$ $\omega(1650)$, $\phi(???)$ \rule{0pt}{3.7ex}%
\rule[-2ex]{0pt}{0pt}\\\hline
\end{tabular}%
\caption{Summary of the quark-antiquark nonets and their properties.}%
\end{table}%

\bigskip%

\begin{table}[h] \centering
\begin{tabular}
[c]{|c|c|c|c|c|}\hline
Chiral multiplet & Current & $U_{R}(3)\times U_{R}(3)$ & $P$ & $C$%
\rule{0pt}{3.7ex}\rule[-2ex]{0pt}{0pt}\\\hline
$\Phi=S+iP$ & $\sqrt{2}\bar{q}_{R,j}q_{L,i}$ & $U_{L}\Phi U_{R}^{\dagger}$ &
$\Phi^{\dag}$ & $\Phi^{t}$ \rule{0pt}{3.7ex}\rule[-2ex]{0pt}{0pt}\\\hline
$R^{\mu}=V^{\mu}-A^{\mu}$ & $\sqrt{2}\bar{q}_{R,j}\gamma^{\mu}q_{R,i}$ &
$U_{R}R^{\mu}U_{R}^{\dagger}$ & $L_{\mu}$ & $L^{t\mu}$\rule{0pt}{3.7ex}%
\rule[-2ex]{0pt}{0pt}\\\hline
$L^{\mu}=V^{\mu}+A^{\mu}$ & $\sqrt{2}\bar{q}_{L,j}\gamma^{\mu}q_{L,i}$ &
$U_{L}R^{\mu}U_{L}^{\dagger}$ & $R_{\mu}$ & $R^{t\mu}$\rule{0pt}{3.7ex}%
\rule[-2ex]{0pt}{0pt}\\\hline
$\tilde{\Phi}^{\mu}=E_{\text{ang}}^{\mu}-iB^{\mu}$ & $\sqrt{2}\bar{q}%
_{R,j}i\overleftrightarrow{\partial}^{\mu}q_{L,i}$ & $U_{L}\tilde{\Phi}^{\mu
}U_{R}^{\dagger}$ & $\tilde{\Phi}^{\dag\mu}$ & $-\tilde{\Phi}^{t\mu}%
$\rule{0pt}{3.7ex}\rule[-2ex]{0pt}{0pt}\\\hline
\end{tabular}%
\caption{Transformation properties of the chiral multiplets.}%
\end{table}%

\section{Decay of the vector glueball into conventional mesons}

In this section we present the interaction terms describing the coupling of
the vector glueball field $\mathcal{O}^{\mu}$ to the various quark-antiquark
multiplets introduced in the previous section. Chiral symmetry as well as
invariance under parity and charge conjugations are the guiding principles. In
addition, dilatation invariance is assumed to hold in the two most relevant
terms. A third interaction which breaks dilatation invariance and involves the
Levi-Civita tensor is also considered. Branching ratios are summarized in the
tables III-VI.

\subsection{Decays into (pseudo)scalar and excited vector and pseudovector
mesons}

The (nontrivial) chiral Lagrangian with the minimal number of quark-antiquark
fields contains the coupling of the vector glueball to (pseudo)scalar and
excited(pseudo)vector mesons:%

\begin{equation}
\mathcal{L}_{1}=\lambda_{\mathcal{O},1}G\mathcal{O}_{\mu}Tr\left[
\Phi^{\dagger}\tilde{\Phi}^{\mu}+\tilde{\Phi}^{\mu\dag}\Phi\right]  \text{ .}
\label{lag1}%
\end{equation}
We introduced also the dilaton $G$ in such a way that the interaction term has
exactly dimension $4$ ($\lambda_{\mathcal{O},1}$ is dimensionless) as required
by dilatation invariance (we recall that only positive or vanishing powers of
$G$ are acceptable \cite{dick}.) Using Table II we obtain:%
\begin{equation}
\mathcal{L}_{1}=\lambda_{\mathcal{O},1}G\mathcal{O}_{\mu}Tr\left[
2SE_{\text{ang}}^{\mu}-2PB^{\mu}\right]  \text{ .}%
\end{equation}
Setting the dilaton field $G$ to its condensate $G=G_{0}$ \cite{schechter},
substituting $P\rightarrow\mathcal{P}$ as described in\ Appendix A, and by
introducing the physical kaonic fields defined in Eqs. (\ref{mixk}), lead to
the desired expressions for the two-body decays (see Appendix B for its
analytic form). In particular, we have decays of the type $\mathcal{O}%
\rightarrow BP$ and $\mathcal{O}\rightarrow SE_{\text{ang}}$. The main decay
channel is $\mathcal{O}\rightarrow b_{1}\pi.$ We thus expect a significant
decay rate of an hypothetical vector glueball into the channel
\begin{equation}
\mathcal{O}\rightarrow b_{1}\pi\rightarrow\omega\pi\pi\text{ .}%
\end{equation}
Unfortunately, we cannot determine $\lambda_{\mathcal{O},1}$ in the present
framework, but we can easily calculate various decay ratios which represent
clear predictions of the present approach, see Table III. Besides the channel
$b_{1}\pi$, also the decays involving kaons and $K_{1}(1270)$ and
$K_{1}(1400)$ are sizable.

In Table III (as well as in all other tables presented in this work) we keep
for definiteness two significant digits for our results. It is difficult to
estimate the actual uncertainty of our ratios, since various unknown factors
enter. One source of error is related to the mass of the glueball, for which
we used the quenched lattice result of 3.81 GeV (one may estimate it $\sim
10$-$15\%$ \cite{mainlattice} see also the detailed discussion in\ Sec. IV.B
and the cautionary remarks written there). Another source of uncertainty is
connected to the validity of the employed effective model; indeed, in Ref.
\cite{walaa} an application of the eLSM to heavy charmed mesons has found to
be applicable within $10\%$ for what concerns the calculation of decays. It
seems reasonable to expect a similar accuracy in the present approach, even if
it is not yet possible to test the decays of heavy glueballs since they were
not yet discovered. Putting altogether, we can estimate the validity of our
branching ratios to about $20$-$30\%.$ However, the interesting point is that
the qualitative outcomes of our study are not strongly dependent on the
precise input of the parameters. For instance, in Table 1 the main information
($O\rightarrow b_{1}\pi$ dominates) is stable.%

\begin{table}[h] \centering
\begin{tabular}
[c]{|c|c|}\hline
Quantity & Value\T\B \rule{0pt}{3.7ex}\rule[-2ex]{0pt}{0pt}\\\hline
$\frac{_{\mathcal{O\rightarrow}\eta h_{1}(1170)}}{_{\mathcal{O\rightarrow
}b_{1}\pi}}$ & 0.17\T\B \rule{0pt}{3.7ex}\rule[-2ex]{0pt}{0pt}\\\hline
$\frac{_{\mathcal{O\rightarrow}\eta h_{1}(1380)}}{_{\mathcal{O\rightarrow
}b_{1}\pi}}$ & 0.11\T\B \rule{0pt}{3.7ex}\rule[-2ex]{0pt}{0pt}\\\hline
$\frac{_{\mathcal{O\rightarrow}\eta^{\prime}h_{1}(1170)}}%
{_{\mathcal{O\rightarrow}b_{1}\pi}}$ & 0.15\T\B \rule{0pt}{3.7ex}\rule[-2ex]%
{0pt}{0pt}\\\hline
$\frac{_{\mathcal{O\rightarrow}\eta^{\prime}h_{1}(1380)}}%
{_{\mathcal{O\rightarrow}b_{1}\pi}}$ & 0.10\T\B\\\hline
$\frac{_{\mathcal{O\rightarrow}KK_{1}(1270)}}{_{\mathcal{O\rightarrow}b_{1}%
\pi}}$ & 0.75\T\B \rule{0pt}{3.7ex}\rule[-2ex]{0pt}{0pt}\\\hline
$\frac{_{\mathcal{O\rightarrow}KK_{1}(1400)}}{_{\mathcal{O\rightarrow}b_{1}%
\pi}}$ & 0.30\T\B \rule{0pt}{3.7ex}\rule[-2ex]{0pt}{0pt}\\\hline
$\frac{_{\mathcal{O\rightarrow}K_{0}^{\ast}(1430)K^{\ast}(1680)}%
}{_{\mathcal{O\rightarrow}b_{1}\pi}}$ & 0.20\T\B \rule{0pt}{3.7ex}\rule[-2ex]%
{0pt}{0pt}\\\hline
$\frac{_{\mathcal{O\rightarrow}a_{0}(1450)\rho(1700)}}{_{\mathcal{O\rightarrow
}b_{1}\pi}}$ & 0.14\T\B \rule{0pt}{3.7ex}\rule[-2ex]{0pt}{0pt}\\\hline
$\frac{_{\mathcal{O\rightarrow}f_{0}(1370)\omega(1650)}}%
{_{\mathcal{O\rightarrow}b_{1}\pi}}$ & 0.034\T\B \rule{0pt}{3.7ex}\rule[-2ex]%
{0pt}{0pt}\\\hline
\end{tabular}%
\caption
{Branching ratios for the decay of the vector glueball into a psuedoscalar-pseudovector pair and into a scalar-excited-vector pair. }%
\end{table}%

At the present state of knowledge, the decays into scalar mesons can only be
approximate. For this reason, we did not `unmix' $\sigma_{N}$ and $\sigma
_{S}.$ In addition, we also expect a three-body decay (when $G$ in Eq.
(\ref{lag1}) is not set to its v.e.v.):
\begin{equation}
\mathcal{O}\rightarrow Gb_{1}\pi\text{ .}%
\end{equation}
Upon identifying $G$ with $f_{0}(1710)$ (the by far dominant contribution in
the eLSM \cite{staninew}), one obtains a very small decay ratio:%
\begin{equation}
\frac{\Gamma_{\mathcal{O\rightarrow}f_{0}(1710)b_{1}\pi}}{\Gamma
_{\mathcal{O\rightarrow}b_{1}\pi}}=3.9\cdot10^{-6}\text{.}%
\end{equation}
There are further three-body interactions contained in $\mathcal{L}_{1},$ but
they are not kinematically allowed.

Moreover, we did not include in the table the ratio $\omega_{\text{ang}%
,S}\equiv\phi(???)$ because the corresponding state was not yet experimentally
found. Yet, assigning it to a yet hypothetical $\phi(1950)$ state, we obtain
$\Gamma_{\mathcal{O\rightarrow}f_{0}(1510)\phi(1950)}/\Gamma
_{\mathcal{O\rightarrow}b_{1}\pi}\simeq0.037.$

We also neglect the mixing $\mathcal{O}_{\mu}Tr[E_{\text{ang}}^{\mu}]$ arising
when the field $S$ condenses (see Appendix A). Namely, the large mass
difference between $\mathcal{O}$ and excited vector mesons assures that such
mixing is negligible.

\subsection{Coupling to (pseudo)scalar and (axial-)vector mesons}

Next, we consider a chirally invariant and dilatation invariant Lagrangians
which couples $\mathcal{O}$ to three quark-antiquark states. It involves both
(pseudo)scalar and (axial-)vector fields:
\begin{equation}
\mathcal{L}_{2}=\lambda_{\mathcal{O},2}\mathcal{O}_{\mu}Tr\left[  L^{\mu}%
\Phi\Phi^{\dagger}+R^{\mu}\Phi^{\dagger}\Phi\right]  \text{ ,}%
\end{equation}
where $\lambda_{\mathcal{O},2}$ is a dimensionless (unknown) coupling
constant. By taking into account the expression listed in Table II, one
obtains terms delivering three-body and two-body decays. The three-body decays
are given by:%
\begin{equation}
\mathcal{L}_{2,\text{three-body}}=\lambda_{\mathcal{O},2}\mathcal{O}_{\mu
}Tr\left[  2V^{\mu}(P^{2}+S^{2})\right]  +2\lambda_{\mathcal{O},2}%
\mathcal{O}_{\mu}Tr\left[  A^{\mu}2i[P,S]\right]  \text{ ,}%
\end{equation}
hence decays of the type $\mathcal{O}\rightarrow VPP$, $\mathcal{O}\rightarrow
VSS$, $\mathcal{O}\rightarrow APS,$ and $\mathcal{O}\rightarrow PPS$ (the
later obtained by the shift $A\rightarrow Zw\partial P,$ see details in
Appendix B) follow. One of the most relevant decay (the second in magnitude)
is
\[
\mathcal{O}\rightarrow\omega\pi\pi\text{ ,}%
\]
which we use as reference for the ratios listed in Table IV (see Appendix C
for the analytic expression). The channel $\mathcal{O}\rightarrow\pi KK^{\ast
}(892)$ turns out to be the largest, followed by $\omega\pi\pi.$ In the last
line of Table IV we have also reported, as an example, a three-body decay into
$a_{0}(1450)a_{0}(1450)\omega,$ which is however very small. This is the case
for all kinematically allowed $VSS$ decays.

Quite interestingly, the most prominent decay of the Lagrangian $\mathcal{L}%
_{1}$ is $\mathcal{O}\rightarrow b_{1}\pi\rightarrow\omega\pi\pi$ (see Table
III), hence it also generates a $\omega\pi\pi$ final state (the state $b_{1}$
has a dominant decay into $\omega\pi$). At a first approximation, one can
write%
\begin{equation}
\Gamma_{\mathcal{O}\rightarrow\omega\pi\pi}^{\text{tot}}\simeq\Gamma
_{\mathcal{O}\rightarrow\omega\pi\pi}^{\text{direct-from }\mathcal{L}_{2}%
}+\Gamma_{\mathcal{O}\rightarrow b_{1}\pi\rightarrow\omega\pi\pi
}^{\text{indirect-from }\mathcal{L}_{1}}\text{ ,}%
\end{equation}
although strictly speaking interferences can appear (usually they are smaller
than 10\%, see the discussion in Ref. \cite{psgproc}; we will neglect such
interferences in the following). Anyway, both $\mathcal{L}_{1}$ and
$\mathcal{L}_{2}$ agree in predicting a strong signal into the final state
$\omega\pi\pi.$ Similarly, the final state into $\pi KK^{\ast}(892)$ is also
relevant, since it comes directly from $\mathcal{L}_{2}$ and indirectly from
$\mathcal{L}_{1}$ via the channels $KK_{1}(1270)\rightarrow K\pi K^{\ast
}(892)$ and $KK_{1}(1400)\rightarrow K\pi K^{\ast}(892)$ (but the decays of
$K_{1}(1270)$ and $K_{1}(1400)$ do not have a single dominating channel
\cite{pdg}).%

\begin{table}[h] \centering
\begin{tabular}
[c]{|c|c|}\hline
Quantity & Value\T\B\rule{0pt}{3.7ex}\rule[-2ex]{0pt}{0pt}\\\hline
$\frac{_{\mathcal{O\rightarrow}KK\rho}}{_{\mathcal{O\rightarrow}\omega\pi\pi}%
}$ & 0.50\T\B\rule{0pt}{3.7ex}\rule[-2ex]{0pt}{0pt}\\\hline
$\frac{_{\mathcal{O\rightarrow}KK\omega}}{_{\mathcal{O\rightarrow}\omega\pi
\pi}}$ & 0.17\T\B\\\hline
$\frac{_{\mathcal{O\rightarrow}KK\phi}}{_{\mathcal{O\rightarrow}\omega\pi\pi}%
}$ & 0.21\T\B\rule{0pt}{3.7ex}\rule[-2ex]{0pt}{0pt}\\\hline
$\frac{_{\mathcal{O\rightarrow}}\pi KK^{\ast}(892)}{_{\mathcal{O\rightarrow
}\omega\pi\pi}}$ & 1.2\T\B\\\hline
$\frac{_{\mathcal{O\rightarrow}\eta\eta\omega}}{_{\mathcal{O\rightarrow}%
\omega\pi\pi}}$ & 0.064\T\B\rule{0pt}{3.7ex}\rule[-2ex]{0pt}{0pt}\\\hline
$\frac{_{\mathcal{O\rightarrow}\eta\eta^{\prime}\omega}}%
{_{\mathcal{O\rightarrow}\omega\pi\pi}}$ & 0.019\T\B\rule{0pt}{3.7ex}%
\rule[-2ex]{0pt}{0pt}\\\hline
$\frac{_{\mathcal{O\rightarrow}\eta^{\prime}\eta^{\prime}\omega}%
}{_{\mathcal{O\rightarrow}\omega\pi\pi}}$ & 0.019\T\B\rule{0pt}{3.7ex}%
\rule[-2ex]{0pt}{0pt}\\\hline
$\frac{_{\mathcal{O\rightarrow}\eta\eta\phi}}{_{\mathcal{O\rightarrow}%
\omega\pi\pi}}$ & 0.039\T\B\rule{0pt}{3.7ex}\rule[-2ex]{0pt}{0pt}\\\hline
$\frac{_{\mathcal{O\rightarrow}\eta\eta^{\prime}\phi}}{_{\mathcal{O\rightarrow
}\omega\pi\pi}}$ & 0.011\T\B\\\hline
$\frac{_{\mathcal{O\rightarrow}\eta^{\prime}\eta^{\prime}\phi}}%
{_{\mathcal{O\rightarrow}\omega\pi\pi}}$ & 0.011\T\B\rule{0pt}{3.7ex}%
\rule[-2ex]{0pt}{0pt}\\\hline
$\frac{_{\mathcal{O\rightarrow}a_{0}(1450)a_{0}(1450)\omega}}%
{_{\mathcal{O\rightarrow}\omega\pi\pi}}$ & 0.00029\T\B\rule{0pt}{3.7ex}%
\rule[-2ex]{0pt}{0pt}\\\hline
\end{tabular}%
\caption
{Branching ratios for the direct three-body decay of the vector glueball into two (pseudo)scalar and one vector meson.}%
\end{table}%
%

\begin{table}[h] \centering
\begin{tabular}
[c]{|c|c|}\hline
Quantity & Value\T\B\rule{0pt}{3.7ex}\rule[-2ex]{0pt}{0pt}\\\hline
$\frac{_{\mathcal{O\rightarrow}a_{0}(1450)\rho}}{_{\mathcal{O\rightarrow
}\omega\pi\pi}}$ & 0.47\T\B\rule{0pt}{3.7ex}\rule[-2ex]{0pt}{0pt}\\\hline
$\frac{_{\mathcal{O\rightarrow}f_{0}(1370)\omega}}{_{\mathcal{O\rightarrow
}\omega\pi\pi}}$ & 0.15\T\B\rule{0pt}{3.7ex}\rule[-2ex]{0pt}{0pt}\\\hline
$\frac{_{\mathcal{O\rightarrow}K_{0}^{\ast}(1430)K^{\ast}(892)}}%
{_{\mathcal{O\rightarrow}\omega\pi\pi}}$ & 0.30\T\B\rule{0pt}{3.7ex}%
\rule[-2ex]{0pt}{0pt}\\\hline
$\frac{_{\mathcal{O\rightarrow}KK}}{_{\mathcal{O\rightarrow}\omega\pi\pi}}$ &
0.018\T\B\rule{0pt}{3.7ex}\rule[-2ex]{0pt}{0pt}\\\hline
\end{tabular}%
\caption
{Branching ratios for the two-body decay of the vector glueball into one (pseudo)scalar and one (axial-)vector mesons and into a kaon-kaon pair.}%
\end{table}%

Two-body decays from $\mathcal{L}_{2}$ are obtained when one of the $\Phi$
condenses, $\Phi\rightarrow\Phi_{0}+\Phi$:
\begin{equation}
\mathcal{L}_{2,\text{two-body}}=\lambda_{\mathcal{O},2}\mathcal{O}_{\mu
}Tr\left[  V^{\mu}2\{\Phi_{0},S)\right]  -\lambda_{\mathcal{O},2}%
\mathcal{O}_{\mu}Tr\left[  A^{\mu}2i[\Phi_{0},P]\right]  \text{ .}
\label{lag2twobody}%
\end{equation}
Then, the decays $\mathcal{O}\rightarrow VS,$ $\mathcal{O}\rightarrow AP,$ and
$\mathcal{O}\rightarrow PP$ follow (expressions in\ Appendix B). The (most
relevant) decay ratios are listed in Table V. The second term in Eq.
(\ref{lag2twobody}) is suppressed because the decay amplitudes are
proportional to the chirally suppressed difference $(\phi_{N}-\sqrt{2}\phi
_{S})$ (this quantity vanishes in the $U_{V}(3)$ limit). Hence both decay
channels $\mathcal{O}\rightarrow AP$ and $\mathcal{O}\rightarrow PP$ are
expected to be very small. In particular, $\mathcal{L}_{2,\text{two-body}}$
contains only interaction terms of the vector glueball with $K_{1,A}K$. Due to
the fact that the $K_{1,A}$ is a mixture of $K_{1}(1270)$ and $K_{1}(1400)$
(Eq. (\ref{mixk})), one would obtain decay rates into $K_{1}(1270)K$ and
$K_{1}(1400)K$, which are already included in Table III. Strictly speaking,
one could describe these decays only once the ratio $\lambda_{\mathcal{O}%
,2}/\lambda_{\mathcal{O},1}$ is known. However, the contribution proportional
to $\lambda_{\mathcal{O},2}$ is suppressed, hence it represents a correction
to the results listed in table III. These decay modes are then omitted from
Table V. The decay into two kaons (also proportional to $(\phi_{N}-\sqrt
{2}\phi_{S})$) is small as expected.

Finally, the decay into $DD$ (the only decay in charmed mesons which is
kinematically allowed) can be obtained by using the extension of the eLSM to
the four-flavor case \cite{walaa}. Due to the fact that chiral symmetry can be
only considered as very approximate when charmed mesons are considered, the
prediction offers only a qualitative result. Anyway, the ratio%
\[
\frac{\Gamma_{\mathcal{O\rightarrow}DD}}{\Gamma_{\mathcal{O\rightarrow}%
\omega\pi\pi}}\approx0.029
\]
shows that the $\bar{D}D$ mode is also expected to be small. This result is
important because it shows that the vector glueball, even if according to
lattice QCD has a mass above the $\bar{D}D$ threshold, decays only rarely in
charmed mesons. The (direct and indirect) $\omega\pi\pi$ and $\pi KK^{\ast
}(892)$ decay modes are expected to by much larger.

\subsection{Coupling to (axial-)vector mesons}%

\begin{table}[h] \centering
\begin{tabular}
[c]{|c|c|}\hline
Quantity & Value\rule{0pt}{3.7ex}\rule[-2ex]{0pt}{0pt}\\\hline
$\frac{_{\mathcal{O\rightarrow}KK^{\ast}(892)}}{_{\mathcal{O\rightarrow}%
\rho\pi}}$ & 1.3\rule{0pt}{3.7ex}\rule[-2ex]{0pt}{0pt}\\\hline
$\frac{_{\mathcal{O\rightarrow}\eta\omega}}{_{\mathcal{O\rightarrow}\rho\pi}}$
& 0.16\rule{0pt}{3.7ex}\rule[-2ex]{0pt}{0pt}\\\hline
$\frac{_{\mathcal{O\rightarrow}\eta^{\prime}\omega}}{_{\mathcal{O\rightarrow
}\rho\pi}}$ & 0.13\rule{0pt}{3.7ex}\rule[-2ex]{0pt}{0pt}\\\hline
$\frac{_{\mathcal{O\rightarrow}\eta\phi}}{_{\mathcal{O\rightarrow}\rho\pi}}$ &
0.21\rule{0pt}{3.7ex}\rule[-2ex]{0pt}{0pt}\\\hline
$\frac{_{\mathcal{O\rightarrow}\eta^{\prime}\phi}}{_{\mathcal{O\rightarrow
}\rho\pi}}$ & 0.18\rule{0pt}{3.7ex}\rule[-2ex]{0pt}{0pt}\\\hline
$\frac{_{\mathcal{O\rightarrow}\rho a_{1}(1230)}}{_{\mathcal{O\rightarrow}%
\rho\pi}}$ & 1.8\rule{0pt}{3.7ex}\rule[-2ex]{0pt}{0pt}\\\hline
$\frac{_{\mathcal{O\rightarrow}}\omega f_{1}(1285)}{_{\mathcal{O\rightarrow
}\rho\pi}}$ & 0.55\rule{0pt}{3.7ex}\rule[-2ex]{0pt}{0pt}\\\hline
$\frac{_{\mathcal{O\rightarrow}}\omega f_{1}(1420)}{_{\mathcal{O\rightarrow
}\rho\pi}}$ & 0.82\rule{0pt}{3.7ex}\rule[-2ex]{0pt}{0pt}\\\hline
\end{tabular}%
\caption
{Branching ratios for the decay of the vector glueball into a vector-pseudoscalar and vector-axial-vector pair. }%
\end{table}%

As a last interaction term we consider an expression which breaks dilatation invariance:%

\begin{equation}
\mathcal{L}_{3}=\alpha\varepsilon_{\mu\nu\rho\sigma}\partial^{\rho}%
\mathcal{O}^{\sigma}Tr\left[  L^{\mu}\Phi R^{\nu}\Phi^{\dagger}\right]  \text{
,}%
\end{equation}
where $\alpha$ has dimension of energy$^{-2}.$ Even if it is expected to lead
to smaller decay rates than the previous two Lagrangians, the presence of an
`anomalous' Levi-Civita tensor may point to a non-negligible interaction
strength even in presence of an explicit breaking of dilatation symmetry.

We restrict our study to the case in which the scalar fields condenses. Hence,
upon setting $\Phi=\Phi_{0},$ we obtain:%
\begin{equation}
\mathcal{L}_{3,\text{two-body}}=2\alpha\varepsilon_{\mu\nu\rho\sigma}%
\partial^{\rho}\mathcal{O}^{\sigma}Tr\left[  A^{\mu}\Phi_{0}V^{\nu}\Phi
_{0}\right]
\end{equation}
which leads to $O\rightarrow AV$ and $O\rightarrow PV$ (upon shifting $A$)$.$
We chose the decay channel $\mathcal{O}\rightarrow\rho\pi$ as our reference
decay mode (see Appendix B for its expression). In Table VI we present the
branching ratios which follow from $\mathcal{L}_{3}.$The dominant decay modes
are and $\rho\pi$, $KK^{\ast}(892)$, and $\rho a_{1}(1230)$ (with increasing
strength). The $\rho\pi$ and $KK^{\ast}(892)$ modes are also important in the
description of the $\rho\pi$ puzzle of $\psi(2S)$ described in the next Section.

\section{Discussions}

In this section we discuss in detail some important issues related to our
approach and our results.\ First, we motivate the applicability of our
interaction Lagrangians, second we justify the use of glueball's masses form
lattice QCD as an input, and third we present the so-called $\rho\pi$-puzzle
and its connection to the vector glueball.

\subsection{On the applicability of the interaction Lagrangians}

The eLSM\ is a low-energy chiral model valid up to $1.7$ GeV, while the vector
glueball studied in this work has a mass of about $3.8$ GeV. It should be
stressed that the joint model
\begin{equation}
\mathcal{L}_{\text{eLSM}}+\mathcal{L}_{O\text{-eLSM}}%
\end{equation}
(where $\mathcal{L}_{O\text{-eLSM}}=\mathcal{L}_{1}+\mathcal{L}_{2}%
+\mathcal{L}_{2}$ is the sum of the three interaction terms described in the
previous section) should be regarded as a model suited to calculate
exclusively the decays of the field $\mathcal{O}$. However, one should not use
such a model to calculate, for instance, pion-pion scattering or analogous
quantities up to the energy of about 4 GeV, since the approach is
\textit{clearly} not complete for that purpose.

Approaches which couple one heavy field to light mesons were widely used in
the literature, e.g. Refs. \cite{tichy,escribano,closejpsi} and refs. therein.
The idea behind these approaches can be explained at best with a simple
example: the decay of the scalar charmonium state $\chi_{c0}$ with a mass of
$3.41$ GeV into two light pseudoscalar mesons. The field $\chi_{c0}$ is
flavor-blind and, by requiring that the whole interaction Lagrangian is
invariant under $U(3)_{V}$ flavor transformations, one is led to the toy-model
Lagrangian
\begin{equation}
\mathcal{L}_{toy}=g_{\chi}\chi_{c0}Tr[P^{t}P],
\end{equation}
where $P$ is the $3\times3$ matrix of pseudoscalar mesons (see Sec. II) and
$g_{\chi}$ is an effective coupling constant. Upon expanding, one obtains
$\mathcal{L}_{toy}=g_{\chi}\chi_{c0}\left(  \vec{\pi}^{2}+2K^{+}K^{-}%
+2K^{0}\bar{K}^{0}+\eta^{2}+\eta^{\prime2}\right)  $. By taking into account
phase space, the theoretical decay ratios read $\Gamma_{\chi_{c0\rightarrow
KK}}/\Gamma_{\chi_{c0\rightarrow\pi\pi}}=1.28,$ $\Gamma_{\chi_{c0\rightarrow
\eta\eta}}/\Gamma_{\chi_{c0\rightarrow\pi\pi}}=0.32,$ and $\Gamma
_{\chi_{c0\rightarrow\eta^{\prime}\eta^{\prime}}}/\Gamma_{\chi_{c0\rightarrow
\pi\pi}}=0.28,$ which should be compared to the experimental results of
$1.42\pm0.14,$ $0.35\pm0.04,$ and $0.23\pm0.04$, respectively \cite{pdg}%
.\ Considering that this toy approach is very simple and does not contain any
violation of flavor symmetry (it involves only the dominant term in the
large-$N_{c}$ expansion, see below), it works remarkably well when compared to
data. Quite interestingly, the toy model also predicts that $\Gamma
_{\chi_{c0\rightarrow\pi^{0}\eta}}=\Gamma_{\chi_{c0\rightarrow\pi^{0}%
\eta^{\prime}}}=\Gamma_{\chi_{c0\rightarrow\eta\eta^{\prime}}}=0$ (the first
two results are a consequence of isospin symmetry, naturally included in
$U(3)_{V}$ flavour symmetry). Indeed, none of these decays has been seen in
the experiment and only stringent upper values exist \cite{pdg}.

The discussed toy model for the resonance $\chi_{c0}$ is clearly limited, but
shows an important fact: the decays of an heavy state (such as a charmonium
resonance) into light mesons still respects the underlying symmetries of the
light system (which is simply flavor symmetry in the present example). This is
the case even if the employed toy-model cannot be considered as a full
hadronic model valid up to the mass of charm-anticharm states. Indeed, this
principle has been utilized in a large number of papers, see for example Ref.
\cite{tichy,escribano} for what concerns the decay of the heavy charmonium
state $j/\psi$ into pseudoscalar mesons and Ref. \cite{closejpsi} for what
concerns decays of $j/\psi$ involving scalar mesons.

In the present work, the calculation of the decays of the vector glueball
follows the very same simple idea. Indeed, in decays of charmonia, one has
first a conversion to gluons, which then transform to light mesons.
Intuitively speaking, the glueball is dominated by gluons, a situation which
is similar to the intermediate state of a charmonium decay. Moreover,
glueballs have some features similar to that of ordinary mesons: they are made
of heavy constituent gluons \cite{fischer} and their size does not seem to be
different from that of ordinary quark-antiquark states \cite{size}. Hence,
considerations based on symmetry appear to be a valid starting point to get
some informations on decays ratios such as the ones of the vector glueball
presented in this work. Clearly, only the future experimental discovery of
glueballs and/or advanced lattice calculations (see the next subsection) will
be able to test this basic assumption of effective hadronic models.

In order to be more realistic, in the present work a better and more complete
low-energy model with (pseudo)scalar and (axial-)vector states was used: the
eLSM. As explained previously, this hadronic model is based on chiral symmetry
and dilatation invariance (as well as explicit, anomalous, and spontaneous
breaking of these symmetries), it contains a finite number of terms, and has
been used to describe meson phenomenology up to $1.7$ GeV. (It is interesting
to note that in\ Ref. \cite{walaa}, the eLSM was also applied to charmed
mesons. It was found that the decays of such heavy mesons still approximately
reflect chiral symmetry.) Moreover, we took into account that the heavy field
that we couple to the eLSM, the vector glueball, is not only flavour blind,
but also chirally blind: in this way chiral symmetry (together with its
spontaneous breaking) has an influence on the determination of the decays.
While a direct comparison with data is not yet possible, predictions can be
obtained by the calculation of decay ratios. These predictions are model
dependent and still neglect symmetry braking terms and mixing effects. Yet,
some branching ratios might be useful in future search of the vector glueball.
Moreover,\textbf{ }the same approach can be actually applied to all glueballs
listed in the lattice spectrum of Ref. \cite{mainlattice}, as the example of
the vector glueball or the example of the pseudoscalar glueball show
\cite{psg}.

\subsection{On the glueball masses}

At present, the best theoretical method to calculate glueball's masses is
lattice, since it numerically simulates the Yang-Mills (or the QCD)
Lagrangian, hence it takes into account the nonperturbative nature of
interactions involving gluons. The masses calculated in Ref.
\cite{mainlattice} (whose results are also reported in the `quark model'
summary of the PDG \cite{pdg}) were obtained in the so-called quenched
approximation (no quark fluctuations). However, in Ref. \cite{gregory} an
unquenched calculation has been performed. In the conclusions of Ref.
\cite{gregory} it is written: \textquotedblleft\textit{The most conservative
interpretation of our results is that the masses in terms of lattice
representations are broadly consistent with results from quenched
QCD.}\textquotedblright\ This is indeed a promising result for model builders.
If mass shifts due to unquenching are not too large, one may be -with due
care- optimistic that quark-antiquark admixtures do not spoil the presented
picture. On the other hand, it must be stressed that the study of Ref.
\cite{gregory} has not been repeated yet by other groups [see however the very
recent two-flavour study of light glueballs in Ref. \cite{lastlattice}, where
the masses also do not vary much when quarks are included]. A full study
involving glueballs and their mixing with conventional mesons would be very
useful to advance in the field. This is unfortunately not an easy task. As
stated in\ Refs. \cite{dudek,dudek2}, glueballs on the Lattice become
challenging in full QCD and it is very hard to determine their admixture in
physical resonances. Namely, glueball signals deteriorate fast into noise,
making them very hard to extract.

In connection to hadronic models effective models, it should be however also
stressed that the masses of glueballs which enter the Lagrangian should be the
quenched ones (and not the unquenched). Namely the unquenching can be
performed \textit{within} the hadronic model. This is the case of the scalar
glueball studied in\ Ref. \cite{staninew}. We also recall that lattice is the
best among many different approaches toward the calculations of the spectrum
of glueballs, see Refs. \cite{review,giacosameson2016} for a list of various
results. Even in the original works based on bag models, e.g. Ref.
\cite{jaffebag}, the vector glueball turned out to be quite heavy (about 3-4
GeV). Quite interestingly, AdS/QCD also finds a mass of the vector state
within the same range \cite{tan}. If, as we shall describe below, the width of
the vector glueball turns out to be sufficiently narrow (surely, the ratio
$\Gamma/(M-E_{threshold})$ is expected to be sufficiently smaller than 1),
shifts due to mesonic quantum fluctuations \cite{boglione} are also suppressed
and shall not change drastically the mass.

Yet, in order to study mixing effects within hadronic models, first one needs
to identify some promising candidates. Namely, the mixing is strongly
dependent on the mass difference between nearby bare states. Its study will be
possible (and highly needed) when more information will be available. As we
shall discuss in the next subsection, at least in one case such mixing was
estimated to be rather small.

Summarizing, it is important to say that a change in the mass of the vector
glueball of about $300$ MeV (above or below the value $3.8$ GeV used in this
paper) shall not change the overall picture. At present, the use of the well
established quenched lattice value of Ref. \cite{mainlattice} for the vector
glueball seems the most reasonable choice to start with.

\subsection{On the $\rho\pi$ puzzle\emph{ }}

The decay of the vector glueball $\mathcal{O}\rightarrow\rho\pi$ and
$\mathcal{O}\rightarrow KK^{\ast}(892)$ (which arise from $\mathcal{L}_{3}$,
see Table VI) are interesting in connection to the so-called $\rho\pi$ puzzle
\cite{brodsky,robson,hou}. This puzzle has to do with the experimentally
missing $\rho\pi$ and $KK^{\ast}(892)$ decays of the resonance $\psi(2S).$
This state is (predominately) a charmonium which emerges as a radial
excitation of the famous $j/\psi$ meson. Its mass is $3.6806109$ GeV (quite
close to the mass of the vector glueball evaluated by lattice QCD) and its
decay width is very small: $\Gamma_{\psi(2S)}=0.298$ MeV. Due to the
similarity of $j/\psi$ and $\psi(2S),$ one expects that the ratio
\begin{equation}
\frac{\Gamma_{\psi(2S)\rightarrow\text{a certain light meson channel}}}%
{\Gamma_{j/\psi\rightarrow\text{a certain light meson channel}}}%
\simeq0.14\text{ ,}%
\end{equation}
holds for all light channels. The value $0.14$ (the $14\%$ rule) is the ratio
of the decays into $e^{+}e^{-}$ \cite{besvg}. This rule works pretty well for
various decay channels, but is badly broken for $\rho\pi$ and $KK^{\ast}(892)$
channels, which are clearly seen for $j/\psi$ but, as mentioned, are not seen
for $\psi(2S).$

A\ possibility to solve this puzzle invokes the presence of a nearby vector
glueball. A mixing of a bare charmonium $\bar{c}c$ state and a bare glueball
$\mathcal{O}\equiv ggg$ leads to two physical states:
\begin{equation}
\left(
\begin{array}
[c]{c}%
\psi(2S)\\
\mathcal{O}^{\prime}%
\end{array}
\right)  =\left(
\begin{array}
[c]{cc}%
\cos\theta & \sin\theta\\
-\sin\theta & \cos\theta
\end{array}
\right)  \left(
\begin{array}
[c]{c}%
\bar{c}c\text{ (with }n=2,\text{ }L=0,S=1\text{)}\\
\mathcal{O}\equiv ggg
\end{array}
\right)  \text{ .}%
\end{equation}
Then, within this picture the state $\psi(2S)$ does not correspond to a pure
charmonium, but contains (a small) glueball amount. In Ref. \cite{hou} it is
estimated that $\left\vert \theta\right\vert \lesssim2^{\circ}.$ This is in
agreement with the fact that such a glueball-quarkonium mixing is suppressed
in the large-$N_{c}$ limit and by the fact that the vector glueball contains
(at least) three constituents gluons.

In conclusion, for what concerns the decays $\Gamma_{\mathcal{O}%
\rightarrow\text{light mesons}}$, the estimated small mixing with a $2S$
charmonium state, has a small influence, thus justifying a posteriori the
results presented in this work. However, a precise study of this small mixing
must be left for the future (when and if a vector glueball candidate will be
found in that mass region).

\section{Conclusions and Outlook}

In this work we have presented three chirally invariant effective interaction
terms describing two-body and three-body decays of a not-yet discovered vector
glueball into (pseudo)scalar, (axial-)vector and pseudo(excited-)vector
mesons. While the intensity of the coupling constant cannot be determined, one
can predict, in the context of our model, some decay ratios and thus determine
which decay channels are expected to dominate. Hopefully, our results, even if
model dependent and subject to various uncertainties (validity of the
symmetries used to write the Lagrangians, the value of input's mass of the
vector glueball, and the absence of mixing, see below), can be of some help in
future experimental search of the vector glueball. In particular, we have
found the following outcomes. In\textbf{ }the first two interaction terms
(which are also dilatation invariant and are expected to be dominant) the main
decay channels are $\mathcal{O}\rightarrow b_{1}\pi\rightarrow\omega\pi\pi$
(first term) as well as $\mathcal{O}\rightarrow\omega\pi\pi$ and
$\mathcal{O}\rightarrow\pi KK^{\ast}(892)$ (second term), see Table III, IV,
and V for all results. Interestingly, the first and second terms predict
sizable $\omega\pi\pi$ and $\pi KK^{\ast}(892)$ final states, which according
to our results represent the golden channels for the identification of a
vector glueball's candidate. Our third interaction terms breaks dilatation
invariance but was considered because it predicts decays into one vector and
one pseudoscalar meson, in particular $\mathcal{O}\rightarrow\rho\pi$ and
$\mathcal{O}\rightarrow KK^{\ast}(892)$ (Table VI). In turn, these channel may
help to understand the $\rho\pi$ puzzle\emph{ }of $\psi(2S)$.

The width of the vector glueball (and of glueballs in general) cannot yet be
determined theoretically. According to large-$N_{c}$ arguments
\cite{thooft,witten,coleman,lebed}, a glueball decay into two mesons scales as
$\Gamma_{G\rightarrow M_{1}M_{2}}\propto N_{c}^{-2}$, hence it is more
suppressed than OZI-allowed decays of conventional mesons (such as
$\rho\rightarrow\pi\pi$) which scale $\Gamma_{M\rightarrow M_{1}M_{2}%
}^{\text{OZI-allowed}}\propto N_{c}^{-1},$ but less suppressed than
OZI-forbidden decays (such as $j/\psi\rightarrow$ light mesons
\cite{ozi,martinshaw}), which scale $\Gamma_{M\rightarrow M_{1}M_{2}%
}^{\text{OZI-forbidden}}\propto N_{c}^{-3}.$ Large-$N_{c}$ considerations
represent only a qualitative statement (and could be well broken for the
physical value $N_{c}=3,$ as it is e.g. in the case of the axial anomaly), but
they support the hope that (at least some) glueballs are not too wide, hence
one can be cautiously optimistic that (some) glueballs can be detected in the
future (in particular, at the PANDA experiment \cite{panda}, designed also for
that purpose). Moreover, the vector glueball studied in this work is built out
of (at least) three constituent gluons, for which decay requires also the
complete annihilations of three gluons, hence possibly not too large.

For what concerns the mass of the glueball, we have used 3.81 GeV as
determined in\ Ref. \cite{mainlattice}. Even if the uncertainty is still
large, the picture concerning the main decay channels does not qualitatively
change when varying the input's mass. In addition, unquenching effects do not
seem to completely change the picture of glueballs \cite{gregory}, but future
lattice studies are needed to confirm this result and to quantify deviations.
In this respect, the coupling of glueballs to mesons can be a very interesting
future achievement of lattice QCD. Namely, it will be possible to further
constraint models such as the one described in this work.

The mixing of the glueball with other nearby quarkonium states needs also to
be studied in the future. In one particular case, the mixing of the vector
glueball with the predominantly charmonium state $\psi(2S)$ has been estimated
to be rather small \cite{hou}. At present, the lack of candidates for the
vector glueball make a study of mixing not yet possible (in fact, mixing
strongly depends on the precise value of the masses, which are not yet known).
In this work, as a first, necessary step, we thus aimed to study the main
interaction terms of a bare, unmixed vector glueball. As soon as candidates
will be found, it will be very interesting and exciting to study mixing in
more detail.

All the remarks above shows that there is a lot of room for improvement of our
approach in the future. New lattice results and experimental findings will be
of great help to advance in this difficult but exciting field of QCD. In this
respect, this work represents a first step toward the search of the vector
glueball, whose main goal is the identification of the possible dominating
decay channels.

Other glueball states can be studied by following the same procedure outlined
in this work. For instance, the tensor glueball ($J^{PC}=2^{++}$) is expected
to have a mass of about $2.2$ GeV (it is the second lightest glueball
according to lattice QCD \cite{mainlattice}); a good candidate could be the
very narrow resonance $f_{J}(2220)$ \cite{tensor,burakovsky} (at present the
options are $J=2$ or $4$ \cite{pdg}; further experimental information is
needed). A future study within the eLSM should include -besides the states
included in this work- also tensor mesons and their chiral partners, the
pseudotensor mesons. In addition to the tensor glueball, one has a full tower
of states listed by lattice QCD: pseudotensor glueball, pseudovector glueball,
oddballs (glueball with exotic quantum numbers such as $J^{PC}=0^{+-}$ and
$J^{PC}=2^{+-}$) as well as various glueballs with $J=3$. Various branching
ratios are parameter-free once the mass of the glueball is fixed and offer a
useful information toward the future search of these important (and still
missing) states of QCD.

As a final remark, it must be stressed that the upcoming PANDA experiment
\cite{panda} is tailor-made for the search of glueballs, since almost all
glueballs (with the exceptions of oddballs) can be directly formed in
proton-antiproton fusion processes.

\section*{Acknowledgments}

The authors thank Dirk H. Rischke for useful discussions. F.G. acknowledges
support from the Polish National Science Centre NCN through the OPUS project
nr. 2015/17/B/ST2/01625. S.J. acknowledges support from H-QM and HGS-HIRe.

\appendix

\section{The eLSM}

\label{app1}

The Lagrangian of the eLSM is built by requiring chiral symmetry
($U(3)_{R}\times U(3)_{L}$), dilatation invariance, as well as invariances
under charge conjugation $C$ and parity $P$:
\begin{align}
\mathcal{L}_{mes}  &  =%
\mathcal{L}%
_{dil}+\mathrm{Tr}[(D_{\mu}\Phi)^{\dagger}(D^{\mu}\Phi)]-m_{0}^{2}\left(
\frac{G}{G_{0}}\right)  ^{2}\mathrm{Tr}(\Phi^{\dagger}\Phi)-\lambda
_{1}[\mathrm{Tr}(\Phi^{\dagger}\Phi)]^{2}-\lambda_{2}\mathrm{Tr}(\Phi
^{\dagger}\Phi)^{2}\nonumber\\
&  -\frac{1}{4}\mathrm{Tr}[(L^{\mu\nu})^{2}+(R^{\mu\nu})^{2}]+\mathrm{Tr}%
\left[  \left(  \frac{m_{1}^{2}}{2}\left(  \frac{G}{G_{0}}\right)  ^{2}%
+\Delta\right)  (L_{\mu}^{2}+R_{\mu}^{2})\right]  +\mathrm{Tr}[H(\Phi
+\Phi^{\dagger})]\nonumber\\
&  +c_{1}(\mathrm{det}\Phi-\mathrm{det}\Phi^{\dagger})^{2}+i\frac{g_{2}}%
{2}\{\mathrm{Tr}(L_{\mu\nu}[L^{\mu},L^{\nu}])+\mathrm{Tr}(R_{\mu\nu}[R^{\mu
},R^{\nu}])\}\nonumber\\
&  +\frac{h_{1}}{2}\mathrm{Tr}(\Phi^{\dagger}\Phi)\mathrm{Tr}\left(  L_{\mu
}^{2}+R_{\mu}^{2}\right)  +h_{2}\mathrm{Tr}[\left\vert L_{\mu}\Phi\right\vert
^{2}+\left\vert \Phi R_{\mu}\right\vert ^{2}]\nonumber\\
&  +2h_{3}\mathrm{Tr}(L_{\mu}\Phi R^{\mu}\Phi^{\dagger})\text{ ,}
\label{fulllag}%
\end{align}
where $D^{\mu}\Phi=\partial^{\mu}\Phi-ig_{1}(L^{\mu}\Phi-\Phi R^{\mu})$ is the
covariant derivative and
\begin{equation}%
\mathcal{L}%
_{dil}=\frac{1}{2}(\partial_{\mu}G)^{2}-\frac{1}{4}\frac{m_{G}^{2}}%
{\Lambda^{2}}\left(  G^{4}\ln\left\vert \frac{G}{\Lambda}\right\vert
-\frac{G^{4}}{4}\right)
\end{equation}
the dilaton (i.e. the scalar glueball) Lagrangian, see
Refs.\ \cite{dick,staninew} for details.

Spontaneous breaking of chiral symmetry takes place ($m_{0}^{2}<0$).\ As a
consequence, one has to perform the shift of the scalar-isoscalar fields by
their vacuum expectation values $\phi_{N}$ and $\phi_{S}$:
\begin{equation}
\sigma_{N}\rightarrow\sigma_{N}+\phi_{N}\text{ and }\sigma_{S}\rightarrow
\sigma_{S}+\phi_{S}\text{ .} \label{shift}%
\end{equation}
In matrix form:%
\begin{equation}
S\rightarrow\Phi_{0}+S\text{ with }\Phi_{0}=\frac{1}{\sqrt{2}}\left(
\begin{array}
[c]{ccc}%
\frac{\phi_{N}}{\sqrt{2}} & 0 & 0\\
0 & \frac{\phi_{N}}{\sqrt{2}} & 0\\
0 & 0 & \phi_{S}%
\end{array}
\right)  \text{ .}%
\end{equation}
In addition, one has also to `shift' the axial-vector fields%
\begin{align}
\vec{a}_{1}^{\mu}  &  \rightarrow\vec{a}_{1}^{\mu}+Z_{\pi}w_{\pi}\partial
^{\mu}\vec{\pi}\text{ , }K_{1,A}^{+,\mu}\rightarrow K_{1,A}^{+,\mu}+Z_{K}%
w_{k}\partial^{\mu}K\text{, ...}\nonumber\\
f_{1,N}^{\mu}  &  \rightarrow f_{1,N}^{\mu}+Z_{\eta_{N}}w_{\eta_{N}}%
\partial^{\mu}\eta_{N}\text{ , }f_{1,S}^{\mu}\rightarrow f_{1,S}^{\mu}%
+Z_{\eta_{S}}w_{\eta_{S}}\partial^{\mu}\eta_{S}\text{ ,}%
\end{align}
and to consider the wave-function renormalization of the pseudoscalar fields:%
\begin{align}
\vec{\pi}  &  \rightarrow Z_{\pi}\vec{\pi}\text{ , }K^{+}\rightarrow
Z_{K}K^{+}\text{, ...}\\
\eta_{N}  &  \rightarrow Z_{\eta_{N}}\eta_{N}\;,\eta_{S}\rightarrow
Z_{\eta_{S}}\eta_{S}\text{ .}%
\end{align}
The constants entering into the previous expressions are:%
\begin{equation}
Z_{\pi}=Z_{\eta_{N}}=\frac{m_{a_{1}}}{\sqrt{m_{a_{1}}^{2}-g_{1}^{2}\phi
_{N}^{2}}}\text{ , }Z_{K}=\frac{2m_{K_{1,A}}}{\sqrt{4m_{K_{1,A}}^{2}-g_{1}%
^{2}(\phi_{N}+\sqrt{2}\phi_{S})^{2}}}\text{ , }Z_{\eta_{S}}=\frac{m_{f_{1S}}%
}{\sqrt{m_{f_{1S}}^{2}-2g_{1}^{2}\phi_{S}^{2}}}\;\text{,} \label{zpi}%
\end{equation}
and:%
\begin{equation}
w_{\pi}=w_{\eta_{N}}=\frac{g_{1}\phi_{N}}{m_{a_{1}}^{2}}\;\text{,}\quad
w_{K}=\frac{g_{1}(\phi_{N}+\sqrt{2}\phi_{S})}{2m_{K_{1},A}^{2}}\text{ ,
}w_{\eta_{S}}=\frac{\sqrt{2}g_{1}\phi_{S}}{m_{f_{1S}}^{2}}\;\;\text{.}
\label{wf1}%
\end{equation}
The numerical values of the renormalization constants are $Z_{\pi}=1.709$,
$Z_{K}=1.604,$ $Z_{\eta_{S}}=1.539$ \cite{dick}, while those of the
$w$-parameters are: $w_{\pi}=0.683$ GeV$^{-1},$ $w_{K}=0.611$ GeV$^{-1}$ ,
$w_{\eta_{S}}=0.554$ GeV$^{-1}$. Moreover, the condensates $\phi_{N}$ and
$\phi_{S}$ read%
\begin{equation}
\phi_{N}=Z_{\pi}f_{\pi}=0.158\text{ GeV, }\phi_{S}=\frac{2Z_{K}f_{K}-\phi_{N}%
}{\sqrt{2}}=0.138\text{ GeV}\;\text{,}%
\end{equation}
where the standard values $f_{\pi}=0.0922$ GeV and $f_{K}=0.110$ GeV have been
used \cite{pdg}. The previous expression can be summarized by the matrix
replacements
\begin{equation}
P\rightarrow\mathcal{P}=\frac{1}{\sqrt{2}}\left(
\begin{array}
[c]{ccc}%
\frac{Z_{\pi}}{\sqrt{2}}(\eta_{N}+\pi^{0}) & Z_{\pi}\pi^{+} & Z_{K}K^{+}\\
Z_{\pi}\pi^{-} & \frac{Z_{\pi}}{\sqrt{2}}(\eta_{N}-\pi^{0}) & Z_{K}K^{0}\\
Z_{K}K^{-} & Z_{K}\bar{K}^{0} & Z_{\eta_{S}}\eta_{S}%
\end{array}
\right)  \text{ ,}%
\end{equation}
and%
\begin{equation}
A^{\mu}\rightarrow\mathcal{A}^{\mu}=\frac{1}{\sqrt{2}}\left(
\begin{array}
[c]{ccc}%
\frac{f_{1N}+a_{1}^{0}}{\sqrt{2}} & a_{1}^{+} & K_{1,A}^{+}\\
a_{1}^{-} & \frac{f_{1N}-a_{1}^{0}}{\sqrt{2}} & K_{1,A}^{0}\\
K_{1,A}^{-} & \bar{K}_{1,A}^{0} & f_{1S}%
\end{array}
\right)  ^{\mu}+\frac{\partial^{\mu}}{\sqrt{2}}\left(
\begin{array}
[c]{ccc}%
\frac{Z_{\pi}w_{\pi}}{\sqrt{2}}(\eta_{N}+\pi^{0}) & Z_{\pi}w_{\pi}\pi^{+} &
Z_{K}w_{K}K^{+}\\
Z_{\pi}w_{\pi}\pi^{-} & \frac{Z_{\pi}w_{\pi}}{\sqrt{2}}(\eta_{N}-\pi^{0}) &
Z_{K}w_{K}K^{0}\\
Z_{K}w_{K}K^{-} & Z_{K}w_{K}\bar{K}^{0} & Z_{\eta_{S}}w_{\eta_{S}}\eta_{S}%
\end{array}
\right)  \text{ .}%
\end{equation}
In the $U_{V}(3)$ limit (in which all three bare quark masses are equals), one
has: $\Phi_{N}=\sqrt{2}\Phi_{S}$, $Z=Z_{\pi}=Z_{K}=Z_{\eta_{S}}$, and
$w=w_{\pi}=w_{K}=w_{\eta_{S}}.\ $Hence, in this limit: $P\rightarrow
\mathcal{P}=ZP$ and $A^{\mu}\rightarrow\mathcal{A}^{\mu}=A+Zw\partial^{\mu}P$.

The eLSM has been also enlarged to four flavors in\ Ref. \cite{walaa}.
Interestingly, charmed meson masses and large-$N_{c}$ dominant decays can be
described relatively well (even if one is far from the natural domain of
chiral symmetry).

In the end, we also recall that the pseudoscalar glueball can be coupled to
the eLSM via the chiral Lagrangian $%
\mathcal{L}%
_{\tilde{G}}=ic_{\tilde{G}\Phi}\tilde{G}\left(  \text{\textrm{det}}%
\Phi-\text{\textrm{det}}\Phi^{\dag}\right)  $, which reflects the axial
anomaly in the pseudoscalar-isoscalar sector, see details and results in
Ref.\ \cite{psg,psgproc}. In a recent extension, the very same Lagrangian is
used to study the decay of an hypothetical excited pseudoscalar glueball
\cite{walaaepsg}.

\section{Expressions for two-body decays}

The decay $\mathcal{O}\rightarrow b_{1}\pi$ from $\mathcal{L}_{1}$ reads:%
\begin{equation}
\Gamma_{\mathcal{O}\rightarrow b_{1}\pi}=c_{\mathcal{O}b_{1}\pi}%
\frac{k_{\mathcal{O}b_{1}\pi}}{8\pi M_{\mathcal{O}}^{2}}\left(  \lambda
_{1}G_{0}Z_{\pi}\right)  ^{2}\frac{1}{3}\left(  2+\frac{(M_{\mathcal{O}}%
^{2}-m_{\pi}^{2}+m_{b_{1}}^{2})^{2}}{4M_{\mathcal{O}}^{2}m_{b_{1}}^{2}%
}\right)  \text{ ,}%
\end{equation}
where $c_{\mathcal{O}b_{1}\pi}=3$ is an isospin factor, $M_{\mathcal{O}}=3.8$
GeV is the glueball mass, and $m_{\pi}$ and $m_{b_{1}}$ are the pion and
$b_{1}$ masses. The quantity $k_{\mathcal{O}b_{1}\pi}$ is the modulus of the
three-momentum of one of the two outgoing particles:%
\begin{equation}
k_{\mathcal{O}b_{1}\pi}=\frac{\sqrt{M_{\mathcal{O}}^{4}-2M_{\mathcal{O}}%
^{2}\left(  m_{\pi}^{2}+m_{b_{1}}^{2}\right)  +\left(  m_{\pi}^{2}-m_{b_{1}%
}^{2}\right)  ^{2}}}{2M_{\mathcal{O}}}\text{ .}%
\end{equation}
The decays of the other channels in Tab. III are calculated in an analogous
way, upon taking into account the change of masses, isospin factors, as well
as the constants entering in the amplitudes. The two-body decays of
$\mathcal{L}_{2}$ presented in Table V are calculated by using the same procedure.

We now turn to $\mathcal{L}_{3}.$ The decay $\mathcal{O}\rightarrow\rho\pi$
reads:%
\begin{equation}
\Gamma_{\mathcal{O}\rightarrow\rho\pi}=c_{\mathcal{O}\rho\pi}\frac
{k_{\mathcal{O}\rho\pi}}{8\pi M_{\mathcal{O}}^{2}}\left(  \frac{\alpha}%
{4}w_{\pi}Z_{\pi}\Phi_{N}^{2}\right)  ^{2}\left(  \frac{2}{3}k_{\mathcal{O}%
\rho\pi}^{2}M_{\mathcal{O}}^{2}\right)  \text{ ,}%
\end{equation}
where $c_{\mathcal{O}\rho\pi}=3$ and $k_{\mathcal{O}\rho\pi}$ is the modulus
of the momentum in this case. The other decays $\mathcal{O}\rightarrow VP$ are
calculated in the same way. The last decay that we consider is $\mathcal{O}%
\rightarrow\rho a_{1}(1230)$:%
\begin{equation}
\Gamma_{\mathcal{O}\rightarrow\rho a_{1}(1230)}=c_{\mathcal{O}\rho a_{1}}%
\frac{k_{\mathcal{O}\rho a_{1}}}{8\pi M_{\mathcal{O}}^{2}}\left(  \frac
{\alpha}{4}\phi_{N}^{2}\right)  ^{2}\frac{1}{3}\left(  6M_{\mathcal{O}}%
^{2}+\frac{2k_{\mathcal{O}\rho a_{1}}^{2}M_{\mathcal{O}}^{2}}{m_{\rho}^{2}%
}+\frac{2k_{\mathcal{O}\rho a_{1}}^{2}M_{\mathcal{O}}^{2}}{m_{a_{1}}^{2}%
}\right)
\end{equation}
where $c_{\mathcal{O}\rho a_{1}}=3$ and $k_{\mathcal{O}\rho a_{1}}$ is the
corresponding momentum. Analogous decays in Table VI are calculated in a
similar way.

\section{Three-body decays of $O$ into two pseudoscalar mesons and a vector
meson}

\label{app3}

For completeness we report the explicit expression for the three-body decay
width of the process $\mathcal{O}\rightarrow P_{1}P_{2}V$:%
\begin{equation}
\Gamma_{\mathcal{O}\rightarrow P_{1}P_{2}V}=\frac{s_{\mathcal{O}\rightarrow
P_{1}P_{2}V}}{32(2\pi)^{3}M_{\mathcal{O}}^{3}}\int_{(m_{1}+m_{2})^{2}%
}^{(M_{\mathcal{O}}-m_{3})^{2}}dm_{12}^{2}\int_{(m_{23})_{\min}}%
^{(m_{23})_{\max}}|-i\mathcal{M}_{\mathcal{O}\rightarrow P_{1}P_{2}V}%
|^{2}dm_{23}^{2}\text{ ,}%
\end{equation}
where (see \cite{pdg}):
\begin{align}
(m_{23})_{\min}  &  =(E_{2}^{\ast}+E_{3}^{\ast})^{2}-\left(  \sqrt{E_{2}%
^{\ast2}-m_{2}^{2}}+\sqrt{E_{3}^{\ast2}-m_{3}^{2}}\right)  ^{2}\text{ ,}\\
(m_{23})_{\max}  &  =(E_{2}^{\ast}+E_{3}^{\ast})^{2}-\left(  \sqrt{E_{2}%
^{\ast2}-m_{2}^{2}}-\sqrt{E_{3}^{\ast2}-m_{3}^{2}}\right)  ^{2}\text{ ,}%
\end{align}
and%
\begin{equation}
E_{2}^{\ast}=\frac{m_{12}^{2}-m_{1}^{2}+m_{2}^{2}}{2m_{12}}\text{ , }%
E_{3}^{\ast}=\frac{M_{\mathcal{O}}^{2}-m_{12}^{2}-m_{3}^{2}}{2m_{12}}\text{ .}%
\end{equation}

The quantities $m_{1}$ and $m_{2}$ refer to the masses of the two pseudoscalar
states $P_{1}$ and $P_{2}$, while $m_{3}$ refer to the of an (axial-)vector
state $V.$ We recall also that $m_{ij}^{2}=(k_{i}+k_{j})^{2}$ with $k_{1},$
$k_{2},$ and $k_{3}$ being the four-momenta of the three outgoing particles.
Clearly, $p=k_{1}+k_{2}+k_{3},$ where $p$ is the four-momentum of the vector glueball.

The amplitude $\mathcal{M}_{\mathcal{O}\rightarrow P_{1}P_{2}V}$ is calculated
at tree-level and $s_{\tilde{G}\rightarrow P_{1}P_{2}V}$ is a symmetrization
factor (it equals $1$ if $P_{1}$ and $P_{2}$ are different, it equals $2$ for
$P_{1}=$ $P_{2}$). As an example, we consider the decay into $\pi^{0}\pi
^{0}\omega.$ The amplitude is:
\begin{equation}
|-i\mathcal{M}_{\mathcal{O}\rightarrow\pi^{0}\pi^{0}\omega}|^{2}=\frac
{\lambda_{2}^{2}Z_{\pi}^{4}}{4}\frac{1}{3}\left(  2+\frac{\left(  M^{2}%
+m_{3}^{2}-m_{12}^{2}\right)  ^{2}}{4M_{\mathcal{O}}^{2}m_{3}^{2}}\right)
\text{ ,}%
\end{equation}
and the symmetry factor is $s_{\mathcal{O}\rightarrow\pi^{0}\pi^{0}\omega}=2.$

\end{document}